\documentclass[aps,pra,twocolumn,showpacs,floatfix,superscriptaddress]{revtex4-1}
\usepackage[utf8]{inputenc}
\usepackage[T1]{fontenc}
\usepackage[english]{babel}
\usepackage{float}
\usepackage{color}
\usepackage{graphicx}
\usepackage{amsmath}
\usepackage{amssymb}
\usepackage{indentfirst}
\usepackage{dcolumn}
\usepackage{soul}

\begin{document}
\title{Magnetism of a Co monolayer on Pt(111) capped by overlayers of $5d$ elements: a spin-model study}

\author{E. Simon}
\email{{esimon@phy.bme.hu}}
\affiliation{Department of Theoretical Physics, Budapest University of Technology and Economics, Budafoki \'{u}t 8, H-1111 Budapest, Hungary}
\author{L. R\'{o}zsa}
\affiliation{Department of Physics, University of Hamburg, D-20355 Hamburg, Germany}
%\affiliation{Institute for Solid State Physics and Optics, Wigner Research Centre for Physics, Hungarian Academy of Sciences,
%P.O. Box 49, H-1525 Budapest, Hungary}
\author{K. Palot\'{a}s}
\affiliation{Department of Complex Physical Systems, Institute of Physics, Slovak Academy of Sciences, SK-84511 Bratislava, Slovakia}
\affiliation{MTA-SZTE Reaction Kinetics and Surface Chemistry Research Group, University of Szeged, H-6720 Szeged, Hungary}
\author{L. Szunyogh}
\affiliation{Department of Theoretical Physics, Budapest University of Technology and Economics, Budafoki \'{u}t 8, H-1111 Budapest, Hungary}
\affiliation{MTA-BME Condensed Matter Research Group, Budapest University of Technology and Economics, Budafoki \'{u}t 8, H-1111 Budapest, Hungary}

\begin{abstract}
Using first-principles calculations, we study the magnetic properties of a Co monolayer on a Pt(111) surface with a capping monolayer of selected $5d$ elements (Re, Os, Ir, Pt and Au). First we determine the tensorial exchange interactions and magnetic anisotropies characterizing the Co monolayer for all considered systems. We find a close relationship between the magnetic moment of the Co atoms and the nearest-neighbor isotropic exchange interaction, which is attributed to the electronic hybridization between the Co and the capping layers, in spirit of the Stoner picture of ferromagnetism. The Dzyaloshinskii-Moriya interaction is decreased for all overlayers compared to the uncapped Co/Pt(111) system, while even the sign of the Dzyaloshinskii-Moriya interaction changes in case of the Ir overlayer. We conclude that the variation of the Dzyaloshinskii-Moriya interaction is well correlated with the change of the magnetic anisotropy energy and of the orbital moment anisotropy. The unique influence of the Ir overlayer on the Dzyaloshinskii-Moriya interaction is traced by scaling the strength of the spin--orbit coupling of the Ir atoms in Ir/Co/Pt(111) and by changing the Ir concentration in the Au$_{1-x}$Ir$_{x}$/Co/Pt(111) system. Our spin dynamics simulations indicate that the magnetic ground state of Re/Co/Pt(111) thin film is a spin spiral with a tilted normal vector, while the other systems are ferromagnetic.
\end{abstract}

\maketitle

\section{Introduction}
Owing to promising technological applications, chiral magnetic structures have become the focus of current experimental and theoretical research activities \cite{Parkin,Fert}. Chiral magnetism is essentially related to the breaking of space-inversion symmetry, since in this case spin--orbit coupling (SOC) leads to the appearance of the Dzyaloshinskii-Moriya interaction (DMI) \cite{Dzyaloshinsky, Moriya} that lifts the energy degeneracy between noncollinear magnetic states rotating in opposite directions. Noncollinear chiral magnetic structures stabilized by the DMI, such as spin spirals and magnetic skyrmion lattices, have been explored in crystals with bulk inversion asymmetry such as MnSi \cite{ISHIKAWA1976525, NAKANISHI1980995, Muhlbauer-Science}. Magnetic thin films and multilayers with broken interfacial inversion symmetry represent another class of systems in which chiral magnetic structures can emerge. In these systems, magnetic transition metal thin films are placed on heavy metal (e.g. Pt, Ir, W) substrates supplying strong spin--orbit interaction. For instance, spin spiral ground states were reported for Mn monolayers on W(110) \cite{Bode2007,PhysRevB.91.064402} and on W(001) \cite{PhysRevLett.101.027201}, spin spirals and skyrmions were detected in the Pd/Fe/Ir(111) bilayer system\cite{Romming-Science,Dupe2014, PhysRevB.90.094410}, while in the case of an Fe monolayer on Ir(111) the formation of a spontaneous magnetic nanoskyrmion lattice has been observed \cite{Heinze2011}. Competing ferromagnetic and antiferromagnetic isotropic exchange couplings are also capable of stabilizing noncollinear spin structures in magnetic thin films and nanoislands \cite{Phark2014, PhysRevB.96.140407,Hsuarxiv,PhysRevLett.117.157205}, while the chirality of these structures is still determined by the DMI.

Understanding and controlling the sign and strength of the DMI at metallic interfaces is one of the key tasks in exploring and designing chiral magnetic nanostructures. A large number of experiments has been devoted to the study of the influence of different nonmagnetic elements on the DMI at magnetic/nonmagnetic metal interfaces \cite{Ryu2014, PhysRevB.90.020402,acs.nanolett.5b02732}, also supported by first-principles calculations \cite{PhysRevLett.115.267210}. 
Recently, it was shown that at $3d/5d$ interfaces the trend for the DMI follows Hund's first rule as the number of valence electrons in the magnetic layer is varied \cite{PhysRevLett.117.247202}, while for a Co/Pt bilayer it was studied how the DMI depends on the number of occupied states close to the Fermi energy by resolving the DMI in reciprocal space \cite{PhysRevB.96.024450}. It was also demonstrated that the magnetic ground state of an Fe monolayer on $5d$ metal surfaces  is strongly influenced by the electronic properties of the substrate \cite{PhysRevB.79.094411, Simon2014}. Because of the interplay between large spin--orbit coupling and high spin-polarizability, particular attention has been paid to the influence of the heavy metal Ir on the DMI. This includes the formation of noncollinear spin structures in ultrathin magnetic films on Ir substrates \cite{Heinze2011,Dupe2014,PhysRevB.90.094410} and the insertion of Ir into 
%surface bilayers \cite{PhysRevB.96.094408} and
multilayer structures \cite{Moreau-Luchaire,Soumyanarayanan,Zeissler,ApplPhysLett.111.132403}. It was demonstrated that the insertion of Ir leads to a sign change of the DMI in the Pt/Co/Ir/Pt system \cite{PhysRevB.90.020402, PhysRevB.94.214422},  and it was suggested that the Ir/Co/Pt stacking order in magnetic multilayers can lead to an enhancement of the DMI \cite{PhysRevLett.115.267210,Moreau-Luchaire}.

Motivated by previous experimental and theoretical investigations, in the present work we explore the role of selected monatomic $5d$ (Re, Os, Ir, Pt, Au) overlayers in influencing the magnetic properties of a Co monolayer deposited on Pt(111). We focus on the investigation of how the electronic hybridization with heavy metal capping layers possessing different numbers of valence electrons and different strengths of the spin--orbit interaction influences the magnetic properties of the Co layer.

In Sec.~\ref{sec2}, the parameters of an extended classical Heisenberg model are discussed, where the coupling between the spins is described by tensorial exchange interactions using first-principles electronic structure calculations. It is also explained how these interactions can be converted to effective or micromagnetic parameters. In Sec.~\ref{sec3a}, the modification of the Co magnetic moments and of the nearest-neighbor isotropic exchange coupling between the Co atoms are found to correlate with the change of the electronic states in the Co and the $5d$ overlayers. In Sec.~\ref{sec3b}, the correlations between the DMI, the magnetic anisotropy energy (MAE) and the orbital moment anisotropy are highlighted. In case of the Ir/Co/Pt(111) system, we find that the DMI in the Co monolayer changes sign compared to Co/Pt(111) and the systems with the other capping layers, and we scale the spin--orbit coupling of the Ir layer in order to get a more profound insight into this phenomenon. This study is supplemented by investigating the DMI and the MAE in Au$_{1-x}$Ir$_{x}$/Co/Pt(111) thin films with an alloy overlayer. Finally, in Sec.~\ref{sec4} we determine the magnetic ground state of the Co monolayer on the Pt(111) substrate with different capping layers using spin dynamics simulations. The results are summarized in Sec.~\ref{sec5}.

\section{Computational methods\label{sec2}}

\subsection{Details of \textit{ab initio} calculations\label{sec2a}}

We performed self-consistent electronic structure calculations for X/Co/Pt(111) (X = Re, Os, Ir, Pt, Au) ultrathin films in terms of the relativistic screened Korringa--Kohn--Rostoker method (SKKR) \cite{Szunyogh,Zeller}. For the case of chemically disordered overlayers, we employed the single-site coherent potential approximation (CPA). We used the local spin density approximation as parametrized by Vosko \textit{et al.} \cite{vosko} and the atomic sphere approximation with an angular momentum cutoff of $\ell_{\textrm{max}}=2$. The energy integrals were performed by sampling 16 points on a semicircle contour in the upper complex energy semiplane. The layered system treated self-consistently consisted of $9$ Pt atomic layers, one Co monolayer, one X monolayer and $4$ layers of vacuum (empty spheres) between the semi-infinite Pt substrate and semi-infinite vacuum. For modeling the geometry of the thin films we used the value $a_{\textrm{2D}}=2.774\,\textrm{\AA}$ for the in-plane lattice constant of the Pt$(111)$ surface and fcc growth was assumed for both the Co and the different overlayers. The distances between the atomic layers were optimized in terms of VASP calculations\cite{Kresse199615, Kresse-PRB, Hafner}. Relative to the interlayer distance in bulk Pt, we found an inward relaxation between $5$\% and $10$\% for the Co monolayer and between $8$\% and $15$\% for the different overlayers.

In order to study the magnetic structure in the Co layer we use the generalized classical Heisenberg model
\begin{equation}
\mathcal{H}=-\frac{1}{2}\sum_{ij}\vec{s}_{i}\mathbf{J}_{ij}\vec{s}_{j}+\sum_{i}\vec{s}_{i}\mathbf{K}\vec{s}_{i},
\label{extHeis}
\end{equation}
where $\vec{s}_{i}$ denotes the spin vector of unit length at site $i$, $\mathbf{J}_{ij}$ is the exchange coupling tensor\cite{rtm-Udvardi}, and $\mathbf{K}$ is the on-site anisotropy matrix. The tensorial exchange coupling can be decomposed into an isotropic, an antisymmetric and a traceless symmetric component \cite{PhysRevB.94.214422},
\begin{equation}
\mathbf{J}_{ij}=J_{ij}\mathbf{I}+\frac{1}{2}\left(\mathbf{J}_{ij}-\mathbf{J}^{T}_{ij}\right)+\left[\frac{1}{2}\left(\mathbf{J}_{ij}+\mathbf{J}^{T}_{ij}\right)-J_{ij}\mathbf{I}\right].
\end{equation}

The isotropic part $J_{ij}=\frac{1}{3}\textrm{Tr}\mathbf{J}_{ij}$ represents the Heisenberg couplings between the magnetic moments. The antisymmetric part of the exchange tensor can be identified with the DM vector,
\begin{equation}
\vec{s}_{i}\frac{1}{2}\left(\mathbf{J}_{ij}-\mathbf{J}^{T}_{ij}\right)\vec{s}_{j}=\vec{D}_{ij}\left(\vec{s}_{i}\times\vec{s}_{j}\right).\label{eqn3}
\end{equation}

From the diagonal elements of the traceless symmetric part of the exchange tensor the two-site anisotropy may be calculated. 

The second term of Eq.~(\ref{extHeis}) comprises the on-site anisotropy with the anisotropy matrix $\mathbf{K}$. Note that for the case of $C_{3v}$ symmetry the studied systems exhibit, the on-site anisotropy matrix can be described by a single parameter, $\vec{s}_{i}\mathbf{K}\vec{s}_{i}=K\left(s_{i}^{z}\right)^{2}$. The effective MAE of the system can be obtained as a sum of the two-site and on-site anisotropy contributions as it will be discussed in Sec.~\ref{sec2c} below. Note that the sign convention for $J_{ij}$, $\boldsymbol{D}_{ij}$ and $K$ are opposite to Ref.~\cite{PhysRevB.94.214422}, from where the values for the Co/Pt(111) system without a capping layer are included for comparison with the present results.
%\begin{equation}
% {\ch K_{\textrm{eff}}=\Delta J+K}.\Delta J=\frac{1}{2}\sum_{j}\left(J_{ij}^{xx}-J_{ij}^{zz}\right)
%\end{equation}

The exchange coupling tensors were determined in terms of the relativistic torque method\cite{rtm-Udvardi, rtm-Ebert}, based on calculating the energy costs due to infinitesimal rotations of the spins at selected sites with respect to the ferromagnetic state oriented along different crystallographic directions. For these orientations we considered the out-of-plane ($z$) direction and three different in-plane nearest-neighbor directions, being sufficient to produce interaction matrices that respect the $C_{3\textrm{v}}$ symmetry of the system. The interaction tensors were determined for all pairs of atoms up to a maximal distance of $5a_{\textrm{2D}}$, for a total of 90 neighbors including symmetrically equivalent ones.

\subsection{Determining the ground state of the system}\label{sec2b}

To find the magnetic ground state of the Co monolayer, we calculated the energies of flat harmonic spin spiral configurations,
\begin{eqnarray}
\vec{s}_{i}=\vec{e}_{1}\cos\vec{k}\vec{R}_{i}+\vec{e}_{2}\sin\vec{k}\vec{R}_{i},\label{eqn1}
\end{eqnarray}
where $\vec{k}$ denotes the spin spiral wave vector, $\vec{e}_{1}$ and $\vec{e}_{2}$ are unit vectors perpendicular to each other, and $\vec{R}_{i}$ is the lattice position of spin $\vec{s}_{i}$. Substituting Eq.~(\ref{eqn1}) into Eq.~(\ref{extHeis}) yields
\begin{eqnarray}
\frac{1}{N}E_{\textrm{SS}}\left(\vec{k},\vec{n}\right)&=&-\frac{1}{2}\sum_{\vec{R}_{ij}}\frac{1}{2}\left(\textrm{Tr}\mathbf{J}_{ij}-\vec{n}\mathbf{J}^{\textrm{symm}}_{ij}\vec{n}\right)\cos\vec{k}\vec{R}_{ij}\nonumber
\\
&&-\frac{1}{2}\sum_{\vec{R}_{ij}}\vec{D}_{ij}\vec{n}\sin\vec{k}\vec{R}_{ij}\nonumber
\\
&&+\frac{1}{2}\left(\textrm{Tr}\mathbf{K}-\vec{n}\mathbf{K}\vec{n}\right),\label{eqn2}
\end{eqnarray}
with $\vec{n}=\vec{e}_{1}\times\vec{e}_{2}$ the normal vector of the spiral, $\vec{R}_{ij}=\vec{R}_{j}-\vec{R}_{i}$, and $\mathbf{J}^{\textrm{symm}}_{ij}=\frac{1}{2}\left(\mathbf{J}_{ij}+\mathbf{J}^{T}_{ij}\right)$. The ground state configuration was approximated by optimizing Eq.~(\ref{eqn2}) with respect to $\vec{k}$ and $\vec{n}$, and comparing it to the energy of the ferromagnetic state,
\begin{eqnarray}
\frac{1}{N}E_{\textrm{FM}}\left(\vec{e}_{\textrm{FM}}\right)=-\frac{1}{2}\sum_{\vec{R}_{ij}}\vec{e}_{\textrm{FM}}\mathbf{J}_{ij}\vec{e}_{\textrm{FM}}+\vec{e}_{\textrm{FM}}\mathbf{K}\vec{e}_{\textrm{FM}},\label{eqn3}
\end{eqnarray}
which was minimized with respect to the ferromagnetic direction $\boldsymbol{e}_{\textrm{FM}}$.

Due to the magnetic anisotropy, actual spin spiral configurations become distorted compared to the harmonic shape defined in Eq.~(\ref{eqn1}). In order to take this effect into account, we further relaxed the configurations obtained above using zero-temperature spin dynamics simulations by numerically solving the Landau--Lifshitz--Gilbert equation \cite{Landau, Gilbert}  
\begin{equation}
\frac{\partial \vec{s}_{i}}{\partial t}=-\frac{\gamma}{1+\alpha^{2}}\vec{s}_{i}\times \vec{B}_{i}^{{ \textrm{eff}}}-\frac{\alpha \gamma}{(1+\alpha^{2})s_{i}}\vec{s}_{i}\times (\vec{s}_{i}\times \vec{B}_{i}^{\textrm{eff}}),
\end{equation}
where $\alpha$ is the Gilbert damping parameter and $\gamma=2\mu_{B}/\hbar$ is the gyromagnetic ratio. The effective field $\vec{B}_{i}^{\textrm{eff}}$ is obtained from the generalized Hamiltonian Eq.~(\ref{extHeis}) as
\begin{equation}
 \vec{B}_{i}^{ \textrm{eff}} = - \frac{1}{m} \frac{\partial \cal{H}}{\partial \vec{s}_i} =  
 \frac{1}{m} \sum_{j (\ne i)}  {\bf J}_{ij}  \vec{s}_j -
 \frac{2}{m}  {\bf K}  \vec{s}_i \, .
\end{equation}
The spin magnetic moment of the Co atom $m$ was determined from the electronic structure calculations. We used a two-dimensional lattice of $128\times128$ sites populated by classical spins with periodic boundary conditions and considered the full tensorial exchange interactions and the on-site anisotropy term when calculating the effective field. In all considered cases we found that the harmonic model provided a good approximation for the wave vector and normal vector of the spin spiral or correctly determined the ferromagnetic ground state. We also performed simulations initialized in random initial configurations to investigate whether noncoplanar configurations can emerge in the systems, but found no indication for such a behavior in the absence of external magnetic field.

\subsection{Effective interaction parameters\label{sec2c}}

In order to allow for a comparison between different \textit{ab initio} calculation methods and experimental results, here we discuss how one can transform between the atomic interaction parameters calculated for many different neighbors used in this paper, and effective nearest-neighbor interactions and parameters in the micromagnetic model.

Complex magnetic textures are often studied in terms of micromagnetic models, where it is assumed that the magnetization direction is varying on a length scale much larger than the lattice constant, and the spins may be characterized by the continuous vector field $\vec{s}(\vec{r})$, the length of which is normalized to 1. In order to describe chiral magnetism, for a magnetic monolayer with $C_{3v}$ point-group symmetry, the energy density is usually expressed as
\begin{equation}
e(\vec{s}) = \mathcal{J} \sum_{\alpha=x,y,z}(\vec{\nabla} s^{\alpha})^2 + \mathcal{D} \, w_D(\vec{s}) - \mathcal{K} (s^z)^2 \, ,
\label{free-energy-mmm}
\end{equation}
with the linear Lifshitz invariant,
\begin{equation}
w_D(\vec{s}) = s^z \partial_x s^x - s^x \partial_x s^z +s^z \partial_y s^y - s^y \partial_y s^z \, .
\end{equation}

The relationship between the micromagnetic parameters $\mathcal{J}$, $\mathcal{D}$, and $\mathcal{K}$ and the atomic parameters in Eq.~(\ref{extHeis}) may be obtained by calculating the energy of the same type of spin configurations. Here we will consider spin spiral states with wave vectors along the $y$ direction,
\begin{eqnarray}
\vec{s}\left(\vec{r}\right)= \vec{e}_{z}\cos {ky}+\vec{e}_{y}\sin {ky},\label{eqn4}
\end{eqnarray}
where the plane of the spiral is spanned by the wave vector direction $\vec{e}_{y}$ and the out-of-plane direction $\vec{e}_{z}$, corresponding to cycloidal spin spirals. In the micromagnetic model, the average energy over the spin spiral reads
\begin{eqnarray}
E_{\textrm{micromagnetic}}=\mathcal{J}V_{a}k^{2}+\mathcal{D}V_{a}k -\frac{1}{2}\mathcal{K}V_{a},\label{eqn5}
\end{eqnarray}
if it is calculated for the atomic volume $V_{a}$. For the atomic model one obtains (cf. Eq.~(\ref{eqn2}))
\begin{eqnarray}
E_{\textrm{atomic}}&=&-\frac{1}{2}\sum_{\vec{R}_{ij}}\frac{1}{2}\left( J_{ij}^{yy}+J_{ij}^{zz}\right)\cos k R_{ij}^{y}\nonumber
\\
&&+\frac{1}{2}\sum_{\vec{R}_{ij}}D_{ij}^{x}\sin kR_{ij}^{y}+\frac{1}{2}K.\label{eqn6}
\end{eqnarray}

Expanding Eq.~(\ref{eqn6}) up to second-order terms in $k$ yields
\begin{eqnarray}
E_{\textrm{atomic}}\approx J_{\rm eff}k^{2}+D_{\rm eff}k+\frac{1}{2}K_{\rm eff}\label{eqn7}
\end{eqnarray} 
apart from a constant shift in energy, with the effective spin-model parameters defined as
\begin{eqnarray}
J_{\rm eff}&=& \frac{1}{4} \sum_j J_{ij} (R^y_{ij})^2,\label{effpars1}
\\
D_{\rm eff}&=& \sum_j D^x_{ij} R^y_{ij},\label{effpars2}
\\
K_{\rm eff}&=& K+\frac{1}{2}\sum_{\vec{R}_{ij}}\left( J_{ij}^{yy}-J_{ij}^{zz}\right).\label{effpars3}
\end{eqnarray}

The effective parameters $J_{\rm eff}$ and $D_{\rm eff}$ are also known as spin stiffness and spiralization, respectively \cite{PhysRevB.94.024403,Freimuth2014}. The relationship between the micromagnetic and the effective parameters is given by
\begin{equation}
\mathcal{J} = \frac{1}{V_a} J_{\rm eff} \, , \; \mathcal{D}=\frac{1}{V_a} D_{\rm eff} \, ,  \; \mathcal{K}= -\frac{1}{V_a} K_{\rm eff}.
\end{equation}

Note that it is possible to define the atomic volume as $V_{a}=\frac{\sqrt{3}}{2}a_{\rm 2D}^2t$ where $\frac{\sqrt{3}}{2}a_{\rm 2D}^2$ is the area of the in-plane unit cell and $t$ is the film thickness. In Ref.~\cite{PhysRevLett.115.267210} the value of $t=n_{\textrm{layer}}\sqrt{\frac{2}{3}}a_{\rm 2D}$ was used with $n_{\textrm{layer}}$ the number of magnetic atomic layers, corresponding to the ideal interlayer distance in an fcc lattice along the $(111)$ direction. However, this approximation becomes problematic when lattice relaxations are taken into account at the surface, since in this description the positions of the centers of the atoms are defined instead of the thickness of the layers. Therefore, we used the expression $V_{a}=\frac{4\pi}{3}R_{\textrm{WS}}^{3}$, where $R_{\textrm{WS}}$ is the radius of the atomic spheres used in the SKKR calculations, with $R_{\textrm{WS}}\approx1.49$\,\AA\:for the considered X/Co/Pt(111) systems.

The cycloidal spin spiral defined in Eq.~(\ref{eqn4}) is called clockwise or right-handed for $k>0$, meaning that when looking at the system from the side with the out-of-plane direction towards the top, the spins are rotating clockwise when moving to the right along the modulation direction of the spiral \cite{PhysRevB.78.140403}. For $k<0$, the spin spiral is called counterclockwise or left-handed. According to Eq.~(\ref{eqn5}), the micromagnetic DMI creates an energy difference between the two rotational senses, with $\mathcal{D}>0$ preferring a counterclockwise rotation. Equation~(\ref{effpars2}) demonstrates that the micromagnetic parameter is connected to the $x$ component of the atomic DM vector for spin spirals with wave vectors along the $y$ direction, or the in-plane component $D_{ij}^{\parallel}$ of the vector for general propagation directions. Note that the magnitude of $D_{ij}^{\parallel}$ is the same for all neighbors that can be transformed into each other via the $C_{3v}$ symmetry of the system, while the sign can be defined based on whether the vectors prefer clockwise or counterclockwise rotation of the spins. Note that in the case of $C_{3v}$ symmetry $\vec{D}_{ij}$ also has a nonvanishing $z$-component, the effect of which on domain walls was investigated in Ref.~\cite{PhysRevB.94.214422}.

Finally, we also define nearest-neighbor atomic interaction parameters $J$ and $D$, which reproduce the effective parameters in Eq.~(\ref{effpars1}) and (\ref{effpars2}),
\begin{equation}
J_{\rm eff}=  \frac{3}{4} a_{\rm 2D}^2 J \, ,  \quad  D_{\rm eff}= \frac{3}{2} a_{\rm 2D} D \, ,
\end{equation}
where $D$ is the in-plane component of the nearest-neighbor DM vector with the sign convention discussed above.

%During the calculations we observed that directly calculating the sums in Eqs.~(\ref{effpars1})-(\ref{effpars2}) leads to parameters that reflect the equilibrium properties of the systems poorly. This can be explained by the fact that the expansion at low wave vectors in Eq.~(\ref{eqn7}) is only valid in a very small vicinity of the $k=0$ point if we consider interactions between atoms located far away from each other, and significant numerical difficulties arise when the \textit{ab initio} calculations for determining the spin-model parameters have to be performed with such a high resolution in the Brillouin zone.

Instead of performing the direct summations in Eqs.~(\ref{effpars1})-(\ref{effpars2}), we fitted the spin spiral dispersion relation in Eq.~(\ref{eqn6}) calculated from all interaction parameters in Eq.~(\ref{extHeis}) with an effective nearest-neighbor model containing $J$, $D$, and $K_{\textrm{eff}}$. The fitting was performed in a range that is sufficiently large to avoid numerical problems, but sufficiently small that the micromagnetic approximations may still be considered valid, corresponding to $\left|k\right|a_{\textrm{2D}}/2\pi\le0.1$. We note that this procedure is similar to how the atomic interaction parameters are determined from spin spiral dispersion relations directly obtained from total energy calculations, see e.g. Refs.~\cite{Bode2007,Dupe2014}, but we used the spin model containing interaction parameters between many neighbors to determine the dispersion relation in the first place. We confirmed with spin dynamics simulations that in ferromagnetic systems the domain wall profiles calculated with the full model Hamiltonian (\ref{extHeis}) agree well with the profiles that can be calculated analytically from a micromagnetic model with the interaction parameters obtained using the above procedure. Nevertheless, we found that not all systems can be sufficiently described by the three parameters used in the micromagnetic model, and this discrepancy can be attributed to the competition between ferromagnetic and antiferromagnetic isotropic Heisenberg interactions -- see Sec.~\ref{sec4} for details.

\begin{table}[ht!]
\centering
\begin{ruledtabular}
\begin{tabular}{l r r r}
 & $D$ (meV) & $D_{\rm eff}$ (meV$\cdot$ \AA) & $\mathcal{D}$ (mJ/m$^{2}$)   \\
\hline
Ref.~\cite{PhysRevB.94.214422} & 2.86     &  11.90                                &       15.11                                      \\
Ref.~\cite{Freimuth2014} &  2.72    & 11.30                              &      14.35                        \\
Ref.~\cite{Dupe2014} &     3.60  &  14.98                                  &      19.02                             \\
Ref.~\cite{PhysRevLett.115.267210} &     3.12   &  12.98                                 &      16.48                                        \\
\end{tabular}
\end{ruledtabular}
\caption{Nearest-neighbor atomic ($D$), effective ($D_{\rm eff}$) and micromagnetic ($\mathcal{D}$) DM coupling obtained in several earlier publications for the Co monolayer on Pt(111). Positive values indicate that the counterclockwise (left-handed) chirality is preferred in the system. For a consistent transformation between the different parameters we used the values $a_{\textrm{2D}}=2.774\,\textrm{\AA}$ and $R_{\textrm{WS}}=1.44\,\textrm{\AA}$. For Ref.~\cite{Dupe2014} we took into account the different definition of the atomic interaction parameters compared to Eq.~(\ref{extHeis}). For Ref.~\cite{PhysRevLett.115.267210} we considered the DMI value for the Co(3)/Pt(3) structure and the correction in Ref.~\cite{PhysRevLett.118.219901} .}
\label{table1}
\end{table}
In order to support the comparison of our calculated parameters with corresponding values obtained from experiments or other theoretical approaches we shall present the microscopic, effective and nearest-neighbor atomic parameters as defined above for all considered systems. As an example, in Table~\ref{table1} we present the comparison between DMI values obtained for the Co monolayer on Pt(111) without a capping layer using different \textit{ab initio} calculation methods in Refs.~\cite{PhysRevLett.115.267210,PhysRevB.94.214422,Freimuth2014,Dupe2014}, similarly to the summary given in Ref.~\cite{PhysRevB.96.060410}. Using the above definitions, we find reasonable agreement between the different theoretical descriptions, and all parameters fall into the range where a ferromagnetic ground state is expected based on the experimental investigations in Ref.~\cite{PhysRevB.96.060410}.

\section{Results\label{sec3}}

\subsection{Isotropic exchange interactions \label{sec3a}}

\begin{figure}[b!]
\centering
\includegraphics[width=1.0\columnwidth]{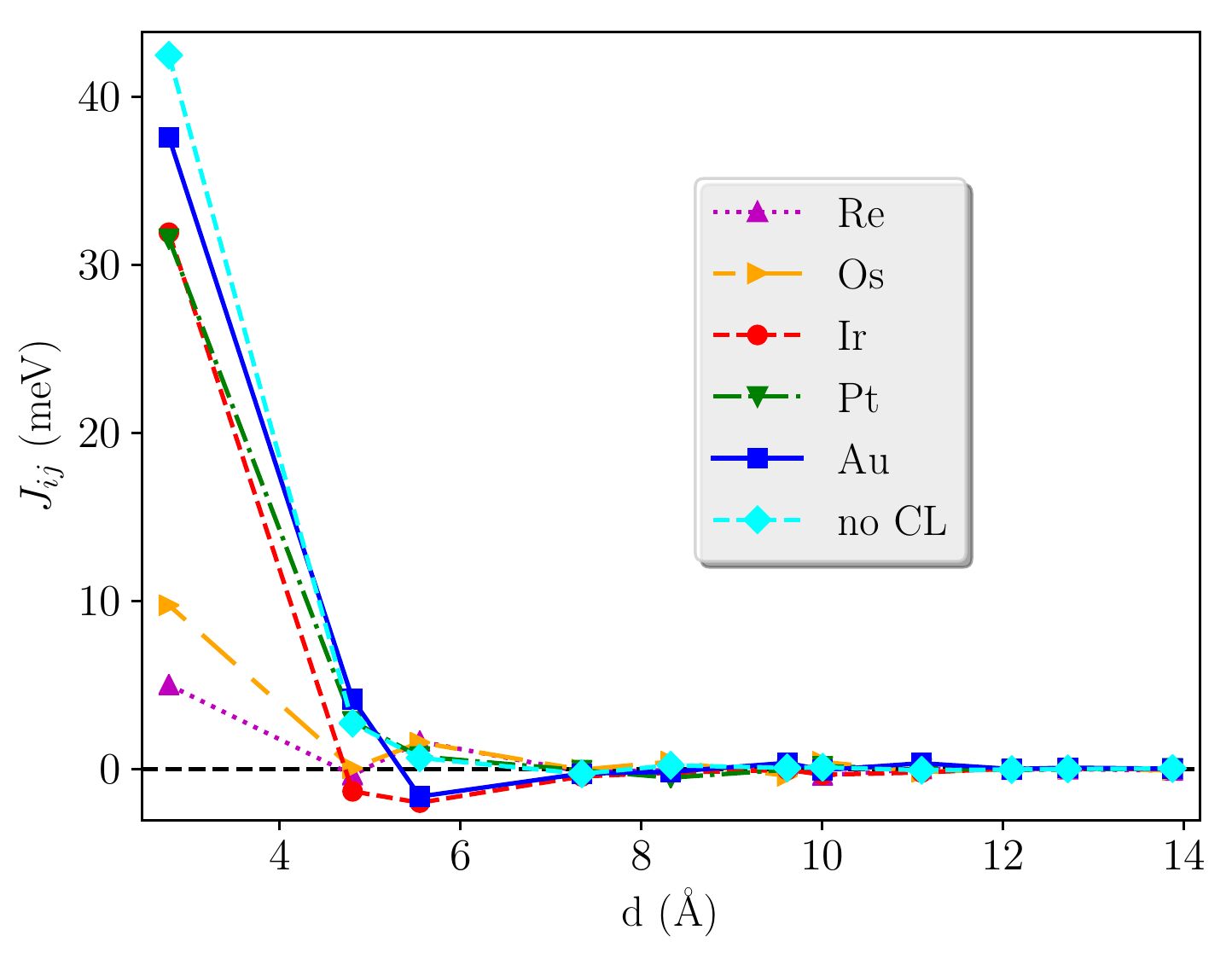}
\vskip -10pt
\caption{(Color online) Calculated Co-Co isotropic exchange parameters $J_{ij}$ as a function of the interatomic distance and different overlayers, and for the Co/Pt(111) system without the capping layer (no CL)\cite{PhysRevB.94.214422}.}
\label{jiso}
\end{figure}

Figure~\ref{jiso} shows the calculated isotropic exchange constants $J_{ij}$ between the Co atoms as a function of interatomic distance for the different overlayers and for the uncapped system (no CL). According to Eq.~(\ref{extHeis}), positive and negative signs of the isotropic exchange parameters refer to ferromagnetic (FM) and antiferromagnetic (AFM) couplings, respectively. For all overlayers the ferromagnetic nearest-neighbor (NN) interaction is dominating: it is the largest in magnitude for the Au overlayer, for Pt and Ir a small decrease can be seen, while for Os and Re overlayers it is dramatically reduced. The second- and third-nearest-neighbor couplings are considerably smaller in magnitude than the NN couplings and the trend for the different overlayers is also less systematic, e.g. in case of Au, Pt and Os overlayers the second-NN coupling is ferromagnetic, while for Ir and Re it is AFM. 
Overall, the magnitude of the isotropic interactions decays rapidly with the distance, becoming negligible beyond the third-NN shell.
\begin{table}[t!]
\centering
\begin{ruledtabular}
\begin{tabular}{l r r r}
 & $J_{1}$ (meV) & $m_{\textrm{Co}}$ ($\mu_{B}$)  \\
\hline
Re &        5.03                               &      1.04                                     \\
Os &        9.66                               &     1.55                        \\
Ir &         31.73                              &     2.11                             \\
Pt &         31.55                             &     2.17                                        \\
Au &        37.54                             &    2.10                                     \\
no CL &    42.46                           &      2.10                                                  \\
\end{tabular}
\end{ruledtabular}
\caption{Calculated nearest-neighbor exchange interactions $J_{1}$ between the Co atoms and the spin-magnetic moment of Co $m_{\textrm{Co}}$ for all considered capping layers and for the Co/Pt(111) system without the capping layer (no CL)\cite{PhysRevB.94.214422}.}
\label{table-iso}
\end{table}

In Table~\ref{table-iso}, the NN exchange couplings ($J_{1}$) and the spin-magnetic moments of Co atoms ($m_{\textrm{Co}}$) are summarized for the different  overlayers. 

We find that capping by $5d$ overlayers systematically reduces $J_{1}$ compared to the uncapped case, which can be attributed to the hybridization between the Co and the overlayer. The magnetic moment of Co is almost constant for the Au, Pt and Ir overlayers, while it shows an apparent decrease for Os and Re, similarly to the NN isotropic exchange. This decrease is, however, much less drastic than for $J_{1}$: $m_{\textrm{Co}}$ in the case of the Re overlayer is about half of $m_{\textrm{Co}}$ in the case of the Au layer, while this ratio is about 1/7 for $J_{1}$. 

\begin{figure}[b!]
\centering
\includegraphics[width=1.0\columnwidth]{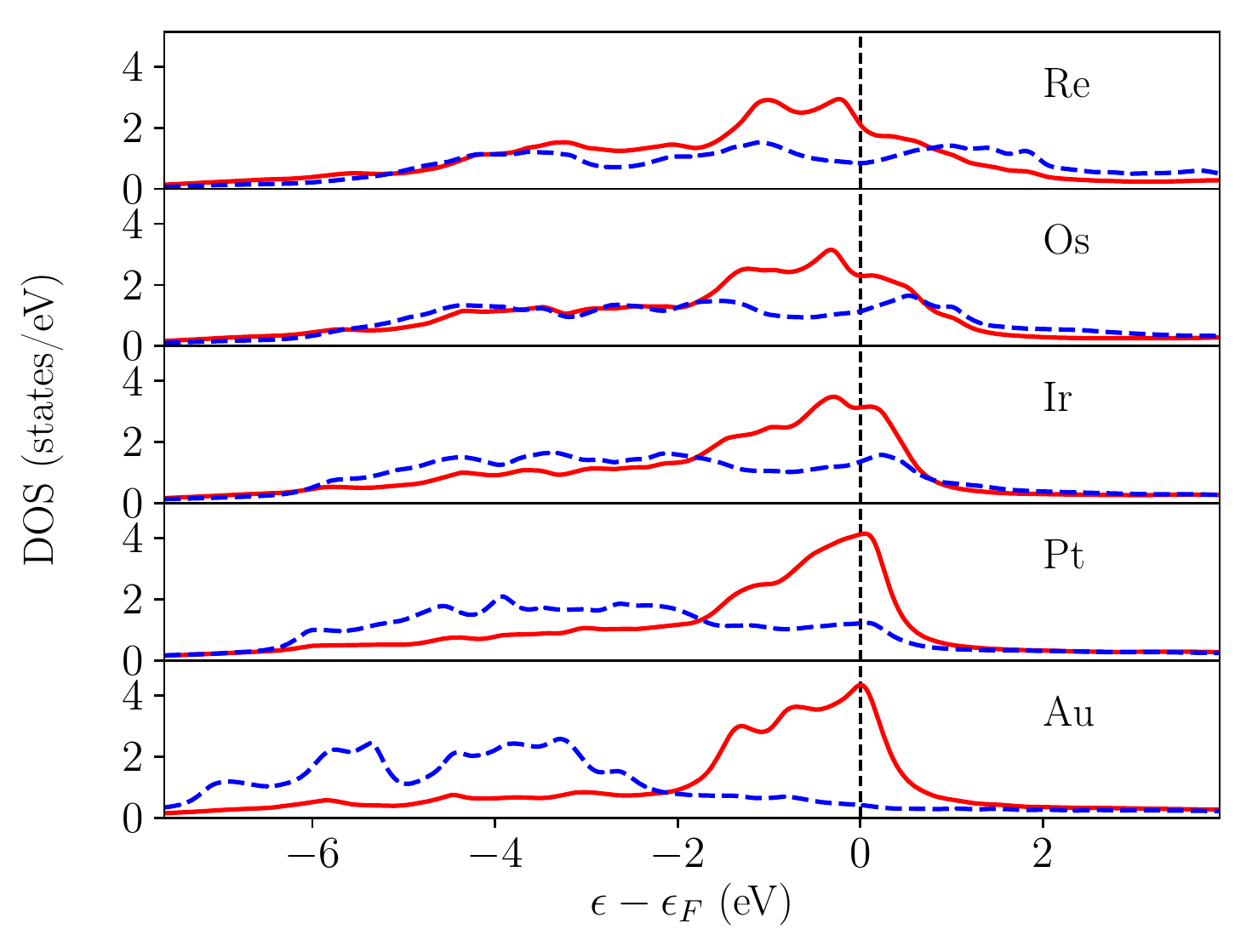}
\vskip -10 pt
\caption{(Color online) DOS of $d$-electrons in the Co layer (solid red line) and in the overlayer (dashed blue line) in non-magnetic X/Co/Pt(111) (X = Re, Os, Ir, Pt, Au) systems.}
\label{dos}
\end{figure}
According to the Stoner model of ferromagnetism, the density of states (DOS) of the $d$-electrons of Co at the Fermi level, $n(\epsilon_F)$, in the nonmagnetic phase plays the crucial role in stabilizing spontaneous magnetization: in case of $I n(\epsilon_F) > 1$ (with $I$ being the Stoner parameter)  the system becomes ferromagnetic.  
Hence the observed trends in $m_{\textrm{Co}}$ and $J_1$ are governed by the filling of the $5d$-band of the overlayer that influences the $3d$-band of Co via hybridization.
In order to trace this effect, in Fig.~\ref{dos} we plot the density of states of the $d$-electrons in the Co layer and in the overlayer in the nonmagnetic phase, meaning that the exchange-correlation magnetic field was set to zero during the density functional theory calculations. Since all the $d$-states of Au are occupied, the corresponding $5d$-band lies well below the Fermi level, leaving the Co $3d$-band localized around the Fermi level, with a large $n(\epsilon_F)$ that explains the strong magnetic moment of Co in this case. Although the $5d$-band of Pt is shifted upwards due to the decrease of the band-filling and the hybridization with the Co $d$-band increases, the large peak in the Co DOS at the Fermi level still pertains, keeping $m_{\textrm{Co}}$ at a high value. This trend remains also in the case of the Ir overlayer, where the $3d$ – $5d$ hybridization further increases and $n(\epsilon_F)$ of Co clearly decreases, but the magnetic moment of Co is of similar value as for the Au overlayer. For the cases of Os and Re overlayers the Co $3d$-band gets rather delocalized due to hybridization with the wider $5d$-bands and $n(\epsilon_F)$ is further reduced leading to the observed drop in $m_{\textrm{Co}}$. Note that a similar dependence of the Co moments on the overlayer was obtained for other systems \cite{PhysRevB.57.8838, PhysRevB.77.184420, Poulopoulos1999, Vaz2008}.

From the calculated isotropic exchange interactions we obtained the spin stiffness constant ($J_{\rm eff}$), the corresponding micromagnetic parameter ($\mathcal{J}$), and NN atomic value ($J$) for all considered overlayers as described in Sec.~\ref{sec2c}, and presented them in Table~\ref{table-jeff}. Apparently, these values follow the variation of $m_{\rm Co}$ or $J_1$ for Os, Pt and Au capping layers; however, in the case of Ir and Re they are considerably reduced. The reason for this behavior is the amplification of the role of exchange interactions between farther atoms in $J_{\rm eff}$ as follows from Eq.~(\ref{effpars1}). From Fig.~\ref{jiso} one can see that in the case of the Ir overlayer both the second- and third-NN couplings are negative (AFM), which drastically reduces the value of $J_{\textrm{eff}}$. The decrease of the NN coupling is apparently insufficient in itself to explain the very small value of $J_{\rm eff}$ in the case of the Re overlayer.
However, a detailed investigation of Fig.~\ref{jiso} shows that the 7$^{\rm th}$-NN interaction, $J_7=-0.39$ meV, gives a dominating negative contribution to $J_{\rm eff}$ due to the large distance ($d=3.606 \, a_{\rm 2D}$) and the large number (12) of neighbors in this shell.
\begin{table}[ht!]
\centering
\begin{ruledtabular}
\begin{tabular}{l   r r r }
& $J$ (meV) & $J_{\rm eff}$ (meV$\cdot$ \AA$^{2}$) & $\mathcal{J}$ (pJ/m)    \\
\hline
Re &   0.82    &  4.73                              &      0.56                                       \\
Os &   22.58    & 130.32                               &     15.48                          \\
Ir &     6.94    & 40.05                              &     4.71                             \\
Pt &    41.89    &241.76                                 &     27.99                                        \\
Au &    49.23    &284.12                                  &    31.98                                       \\
no CL &  54.40  &  313.96       &                            39.86                                                               \\
\end{tabular}
\end{ruledtabular}
\caption{Nearest-neighbor atomic ($J$), effective ($J_{\rm eff}$) and micromagnetic ($\mathcal{J}$) parameters of Co for the isotropic exchange interaction of X/Co/Pt(111) (X=Re, Os, Ir, Pt, Au) and Co/Pt(111) thin films (no CL)\cite{PhysRevB.94.214422} obtained from the calculated spin-model parameters by the fitting procedure in Sec.~\ref{sec2c}.
}
\label{table-jeff}
\end{table}

\subsection{Relativistic spin-model parameters\label{sec3b}}

\subsubsection{Different capping layers}
\begin{figure}[ht!]
\centering
\includegraphics[width=1.0\columnwidth]{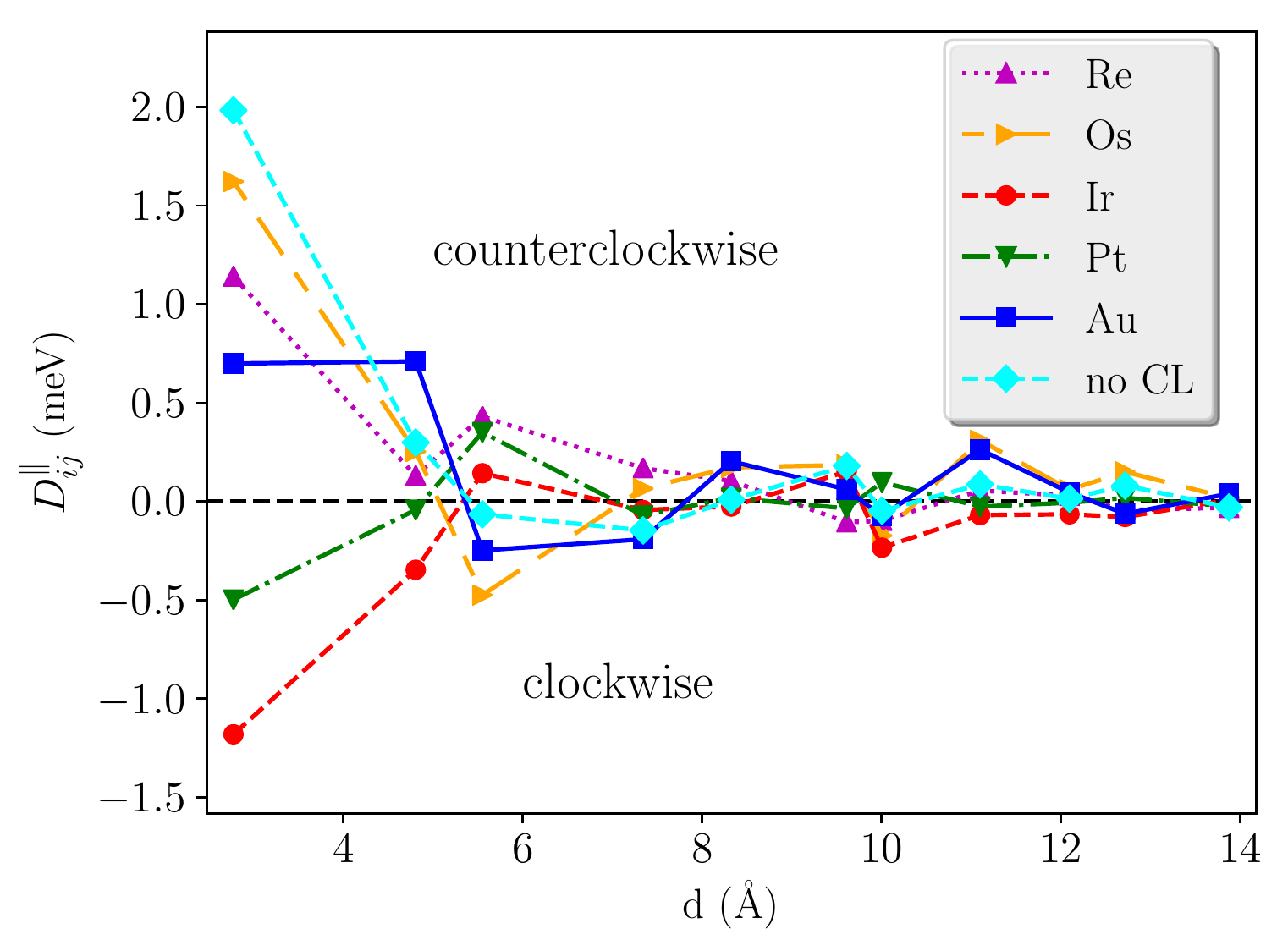}
\vskip -10 pt
\caption{(Color online) In-plane component of the DM vectors $D_{ij}^{\parallel}$ as a function of the distance between the Co atoms for different overlayers, and for the Co/Pt(111) system without the capping layer (no CL)\cite{PhysRevB.94.214422}.}
\label{dm}
\end{figure}

Next, we investigate the in-plane components of Dzyaloshinskii-Moriya interactions between the Co atoms which are shown in Fig.~\ref{dm} for all capping layers as a function of the distance between the Co atoms,  compared to the values in the absence of a capping layer\cite{PhysRevB.94.214422}.
The sign changes of the DMI indicate switchings in the preferred rotational sense from shell to shell, analogously to the oscillation between ferromagnetic and antiferromagnetic isotropic exchange interaction coefficients. Except for the case of the Au capping layer, the NN DMI is the largest in magnitude; however, the DM vectors for more distant pairs also play an important role, since they show a slower decay as a function of the distance compared to the isotropic exchange interactions.
% We recall that in the absence of an overlayer the NN in-plane DMI is $1.98$~meV \cite{PhysRevB.94.214422}.

To illustrate the overall effect of the overlayers on the DMI, we calculated the NN atomic, effective and micromagnetic DMI coefficients of Co from the \textit{ab initio} spin-model parameters as discussed in Sec.~\ref{sec2c}. These values are summarized in Table~\ref{table-deff} for different capping layers. For comparison, we also included the corresponding values for Co/Pt(111). It is worthwhile to mention that the effective parameters in Table~\ref{table-deff} follow exactly the same order for the different capping layers as the in-plane NN DM vectors in Fig.~\ref{dm}, unlike in the case of the isotropic exchange interactions. Regardless of the choice of the capping layer, the DMI is shifted towards the direction of clockwise rotational sense compared to the uncapped system. For the Pt/Co/Pt(111) system, the DMI is exceptionally weak, which is to be expected since inversion symmetry is almost restored in this system if we consider that generally the interfacial DMI is dominated by the magnetic and nonmagnetic heavy metal layers directly next to each other.

We would also like to point out that the Ir capping layer is the only one that switches the sign of the DMI preferring clockwise rotation. This is somewhat unexpected since the Ir layer also changed the preferred rotational sense to clockwise when it was introduced between the Co monolayer and the Pt(111) substrate \cite{PhysRevB.94.214422}, so it should prefer a counterclockwise rotation for the opposite stacking order according to the three-site model of the DMI \cite{PhysRevLett.44.1538}. A possible reason for this effect is that the reduced coordination number of the Ir atoms in the capping layer as well as the electrostatic potential barrier at the surface significantly modify the electronic structure of the capping layer compared to the bulk case or when the Ir is inserted below the Co layer. This sign change of the DMI in Ir/Co/Pt(111) indicates that ultrathin film systems can display qualitatively different features compared to magnetic multilayers, where the Ir/Co/Pt stacking was suggested as a way of enhancing the DMI \cite{Moreau-Luchaire}. The different behavior of Ir as a capping layer and as an inserted layer was recently investigated in Ref.~\cite{PhysRevB.96.094408}.

\begin{table}[t!]
\centering
\begin{ruledtabular}
\begin{tabular}{l r r r}
 & $D$ (meV) & $D_{\rm eff}$ (meV$\cdot$ \AA) & $\mathcal{D}$ (mJ/m$^{2}$)   \\
\hline
Re & 1.82     &  7.57                                &       8.94                                      \\
Os &  2.58    & 10.74                              &      12.75                        \\
Ir &     -1.75  &  -7.28                                  &       -8.56                             \\
Pt &     0.20   &  0.83                                 &      0.96                                        \\
Au &     1.50  &  6.24                                  &     7.02                                       \\
no CL&       2.86   &   11.90                          &     15.11                        \\
\end{tabular}
\end{ruledtabular}
\caption{Nearest-neighbor atomic ($D$), effective ($D_{\rm eff}$) and micromagnetic ($\mathcal{D}$) DM coupling of Co obtained from the spin-model parameters for X/Co/Pt(111) thin films (X=Re, Os, Ir, Pt, Au) and for Co/Pt(111) without any capping layer (no CL).}
\label{table-deff}
\end{table}

\begin{figure}[ht!]
\centering
\includegraphics[width=1.0\columnwidth]{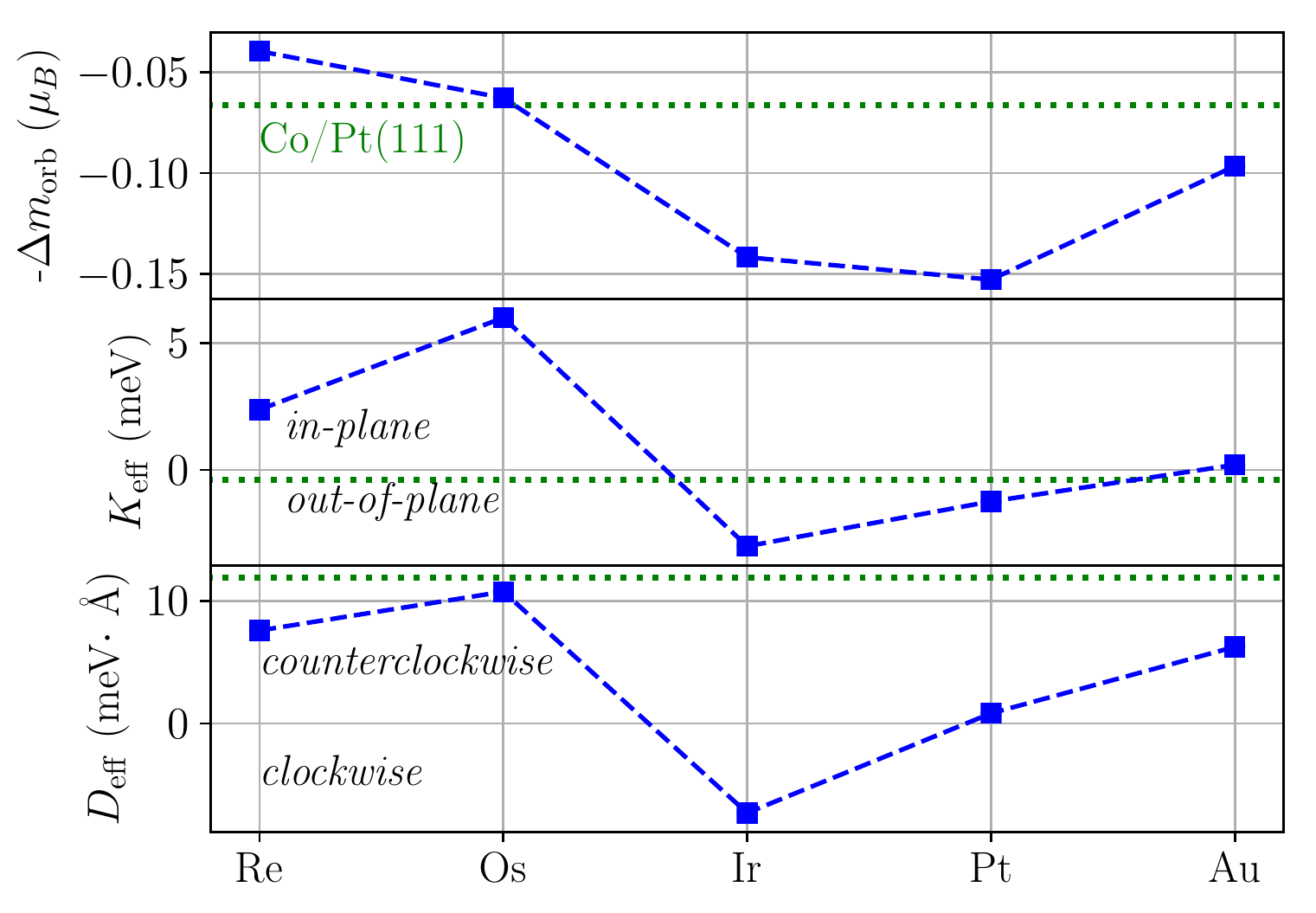}\\
\caption{(Color online) Calculated values of orbital moment anisotropy in the Co layer with negative sign $-\Delta m_{\rm orb}$, MAE $K_{\rm eff}$, and effective DMI $D_{\rm eff}$ for X/Co/Pt(111) thin films (X=Re, Os, Ir, Pt, Au). The corresponding parameters for Co/Pt(111) are also illustrated by dashed green lines. }
\label{eff-dm}
\end{figure}

In order to study the dependence of the DMI on the capping layer, we calculated additional quantities determined by the strength of the spin--orbit coupling, namely the total MAE $K_{\rm eff}$ and the anisotropy of the orbital moment of Co atoms, $\Delta m_{\rm orb}=m_{\rm orb}^{\perp}-m_{\rm orb}^{\parallel}$, where the superscripts $\perp$  and $\parallel$ refer to calculations performed for a normal-to-plane and an in-plane orientation of the magnetization in the Co layer, respectively. Figure~\ref{eff-dm} shows $\Delta m_{\rm orb}$ with a negative sign (top panel), $K_{\rm eff}$  (middle panel), and $D_{\rm eff}$ (bottom panel) for the Co monolayer depending on capping layer. Note that negative and positive signs of $K_{\rm eff}$ refer to easy-axis and easy-plane types of magnetic anisotropy, respectively.

For $3d$ transition metals, where the spin--orbit coupling is small compared to the bandwidth, second-order perturbation theory describes the uniaxial magnetic anisotropy well \cite{PhysRevB.39.865}. According to Bruno's theory, neglecting spin-flop coupling and for a filled spin-majority $d$-band,  a negative proportionality between the MAE and $\Delta m_{\rm orb}$ applies, that was confirmed theoretically and experimentally for Co layers \cite{PhysRevB.39.865, PhysRevLett.77.1805, PhysRevLett.75.3748, PhysRevLett.75.3752, PhysRevB.82.094409}. From Fig.~\ref{eff-dm} a good qualitative correlation can be inferred between $K_{\rm eff}$ and $-\Delta m_{\rm orb}$ with the exception of the Re overlayer. Indeed, due to the large $3d$-$5d$ hybridization, the delocalization of the spin-majority band of Co is increased in the case of the Re overlayer such that the above mentioned conditions for the simple proportionality do not apply.     

From Fig.~\ref{eff-dm} it turns out that the variations of $K_{\rm eff}$ and $D_{\rm eff}$ also correlate well with each other. This is somewhat surprising since, as mentioned above, the MAE is of second order in the SOC, while the DM term appears in the first order of the perturbative expansion \cite{Moriya}. Compared to the Co/Pt(111) system, the Os capping layer does not modify the DMI significantly, but we observe a strong easy-plane MAE. The Re and the Au capping layers decrease the magnitude of $D_{\textrm{eff}}$, and the preferred magnetization direction is also in-plane. An out-of-plane magnetization was obtained for Ir and Pt capping layers, and as discussed above, the Ir capping layer prefers a clockwise rotation, while in the case of the Pt overlayer the DMI is close to zero.

\subsubsection{Scaling of the spin--orbit coupling in the Ir overlayer}

To gain further insight into the the sign change of the DMI in the Co monolayer with the Ir capping layer, we artificially manipulated the strength of SOC at the Ir atoms. Ebert {\em et al.} introduced a continuous scaling of the SOC via the parameter $\lambda$ within the relativistic KKR formalism \cite{PhysRevB.53.7721}: calculation without scaling ($\lambda = 1$) corresponds to the fully relativistic case, while $\lambda = 0$ can be identified with the so-called scalar-relativistic description. Importantly, in the above formalism the scaling of the SOC can be used selectively for arbitrary atomic cells. We thus applied it to the Ir monolayer, while the SOC at all other sites of the system remained unaffected.

\begin{figure}[t!]
\centering
\includegraphics[width=1.0\columnwidth]{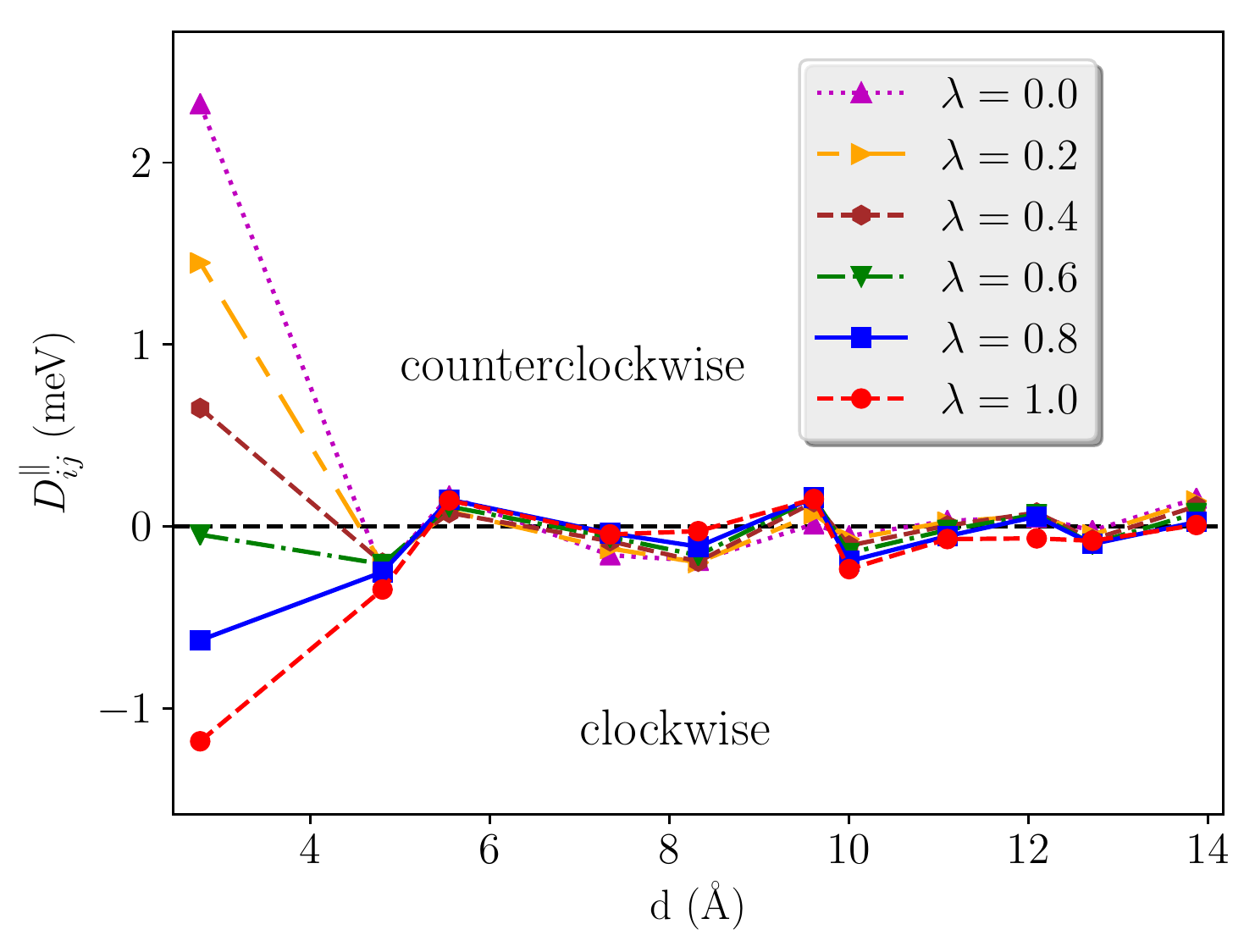}
\vskip -10pt
\caption{(Color online) In-plane DMI  as a function of the distance between the Co atoms for various values of the SOC scaling parameter $\lambda$ in the Ir capping layer of the Ir/Co/Pt(111) system.}
\label{dm-scaling}
\end{figure}

Figure~\ref{dm-scaling} shows $D_{ij}^{\parallel}$ as a function of the distance between the Co atoms for different scaling parameters. Varying $\lambda$ has a strong influence on the NN in-plane DMI: it changes continuously from preferring counterclockwise ($\lambda = 0$) to preferring clockwise ($\lambda = 1$) rotational direction, while the changes in the other shells are smaller in relative and in absolute terms. In the case of $\lambda = 0$, the NN in-plane DMI takes a value of $2.32$ meV,  which means that the NN DMI of the Co/Pt(111) system ($1.98$ meV) \cite{PhysRevB.94.214422} is nearly restored in this case. 

\begin{figure}[htb!]
\centering
\includegraphics[width=1.0\columnwidth]{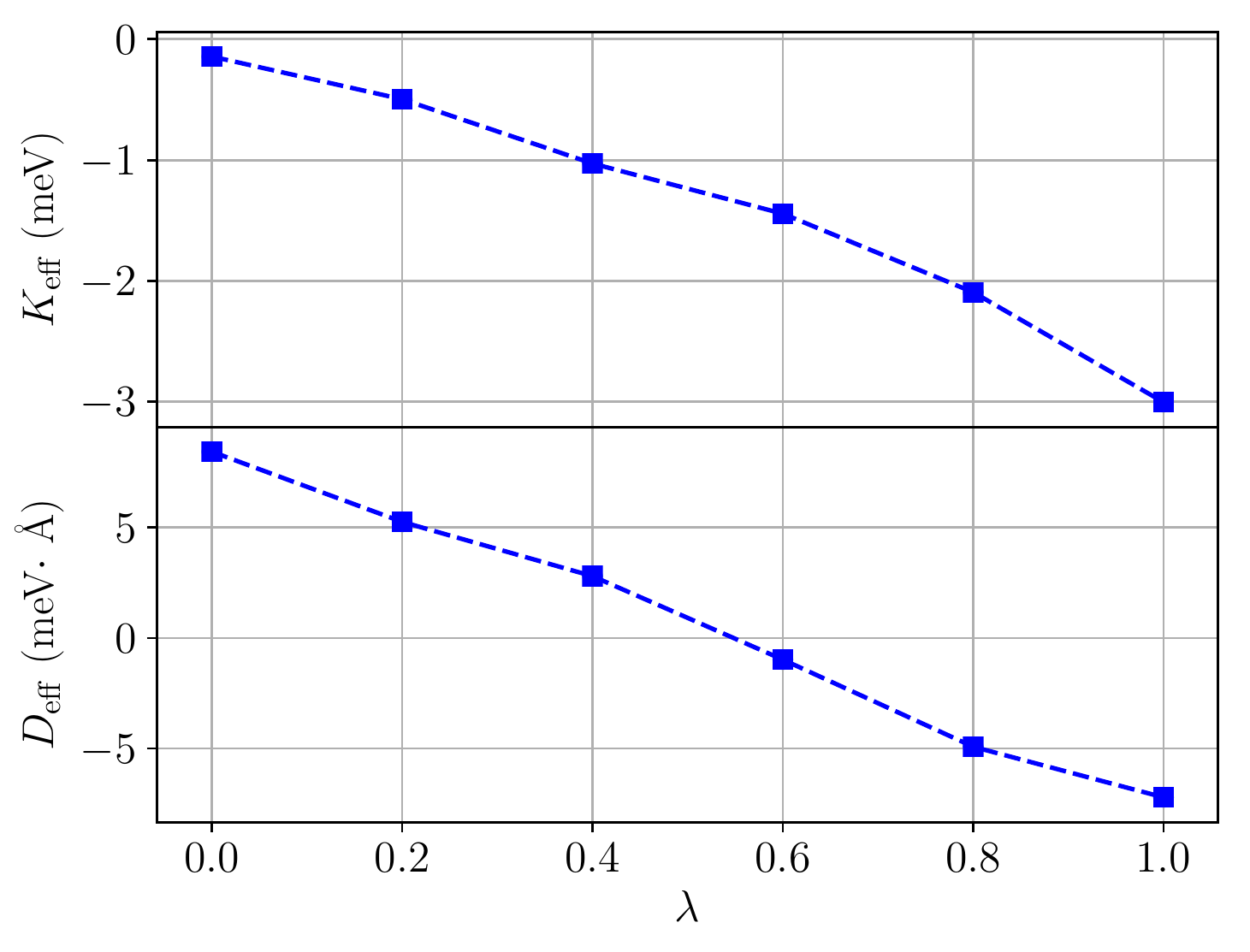}
\vskip -10pt
\caption{(Color online) Calculated MAE $K_{\rm eff}$, and effective DMI $D_{\rm eff}$ as a function of the SOC scaling parameter $\lambda$ in the Ir overlayer of the Ir/Co/Pt(111) system.}
\label{eff-scaling}
\end{figure}

In accordance with the results of first-order perturbation theory, Fig.~\ref{eff-scaling} illustrates that 
the variation of the effective DMI is rather linear with $\lambda$. For $\lambda = 0$, $K_{\rm eff}$ is close to the value of the uncapped Co/Pt(111) system ($-0.20$ meV \cite{PhysRevB.94.214422}) and it increases in magnitude to $-3$ meV for $\lambda = 1$. Following the change in the NN in-plane DMI interaction {in Fig.~\ref{dm-scaling}}, the sign of the effective DMI turns from preferring counterclockwise to preferring clockwise rotation when increasing the strength of the SOC in the Ir overlayer. On the other hand, at $\lambda = 0$ $D_{\rm eff}$ is somewhat smaller in magnitude than in the case of the uncapped Co/Pt(111) ($11.90$ meV$\cdot$\AA). This indicates that the Ir overlayer influences the DMI of the system not just due to its strong SOC, but also by modifying the electronic states in the Co monolayer via hybridization. 

\subsubsection{Changing the capping layer composition in Au$_{1-x}$Ir$_{x}$/Co/Pt(111)}

Controlling the Ir concentration $x$ in the alloy capping layer Au$_{1-x}$Ir$_{x}$ ($0 \le x \le 1$) represents a transition where the effect of increasing hybridization between the $3d$-band of Co and the $5d$-band of the capping metal can be traced, as shown in Fig.~\ref{dos}. On the other hand, the strength of the SOC, defined by the operator $\xi\vec{L}\vec{S}$, in Au and Ir is roughly the same ($\xi\approx 600\,\textrm{meV}$), meaning that the alloying is expected to have a different effect than the scaling of the SOC discussed in the previous section. Thus, we performed calculations of the spin-model parameters for $x= 0.1, 0.2, \dots, 0.9$ by using the coherent-potential approximation (CPA) for the chemically disordered overlayer. The layer relaxation was varied as a function of $x$ according to Vegard's law using the calculated layer relaxation of the Au/Co/Pt(111) and Ir/Co/Pt(111) systems.

\begin{figure}[b!]
\centering
\includegraphics[width=1.0\columnwidth]{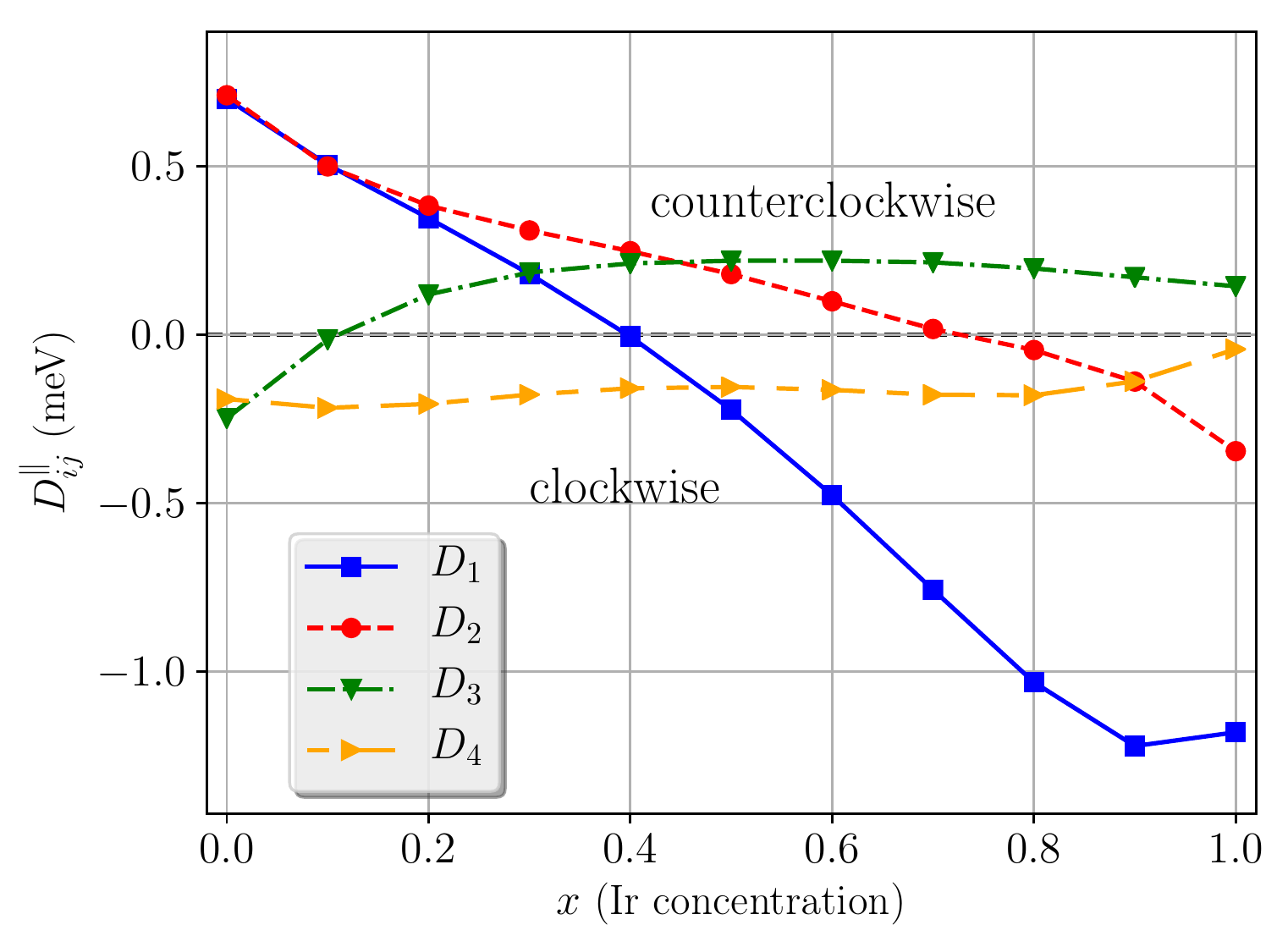}
%\vskip -15pt
\caption{In-plane components of the DM vectors $D_{ij}^{\parallel}$ of Co from the first ($D_{1}$) to the fourth shell ($D_{4}$), as a function of the Ir concentration ($x$) in the Au$_{1-x}$Ir$_{x}$/Co/Pt(111) system. }
\label{fig:dm-alloying}
\end{figure}
The in-plane components of the DM vectors in the Co monolayer from the first to the fourth shell are
shown in Fig.~\ref{fig:dm-alloying} as a function of the Ir concentration. When increasing the Ir concentration, the sign of the first-NN and the second-NN $D_{ij}^{\parallel}$ changes from positive to negative. The third-NN in-plane DM for the Au/Co/Pt(111) system is negative, it turns positive around $x\approx 0.1$, and it has approximately the same magnitude around $20$\% Ir concentration as for the pure Ir/Co/Pt(111) layer, with a maximal amplitude at about $x=0.5$. The sign of the fourth-NN $D_{ij}^{\parallel}$ is not changed by the alloying, and the magnitude remains nearly constant.

\begin{figure}[t!]
\centering
\includegraphics[width=1.0\columnwidth]{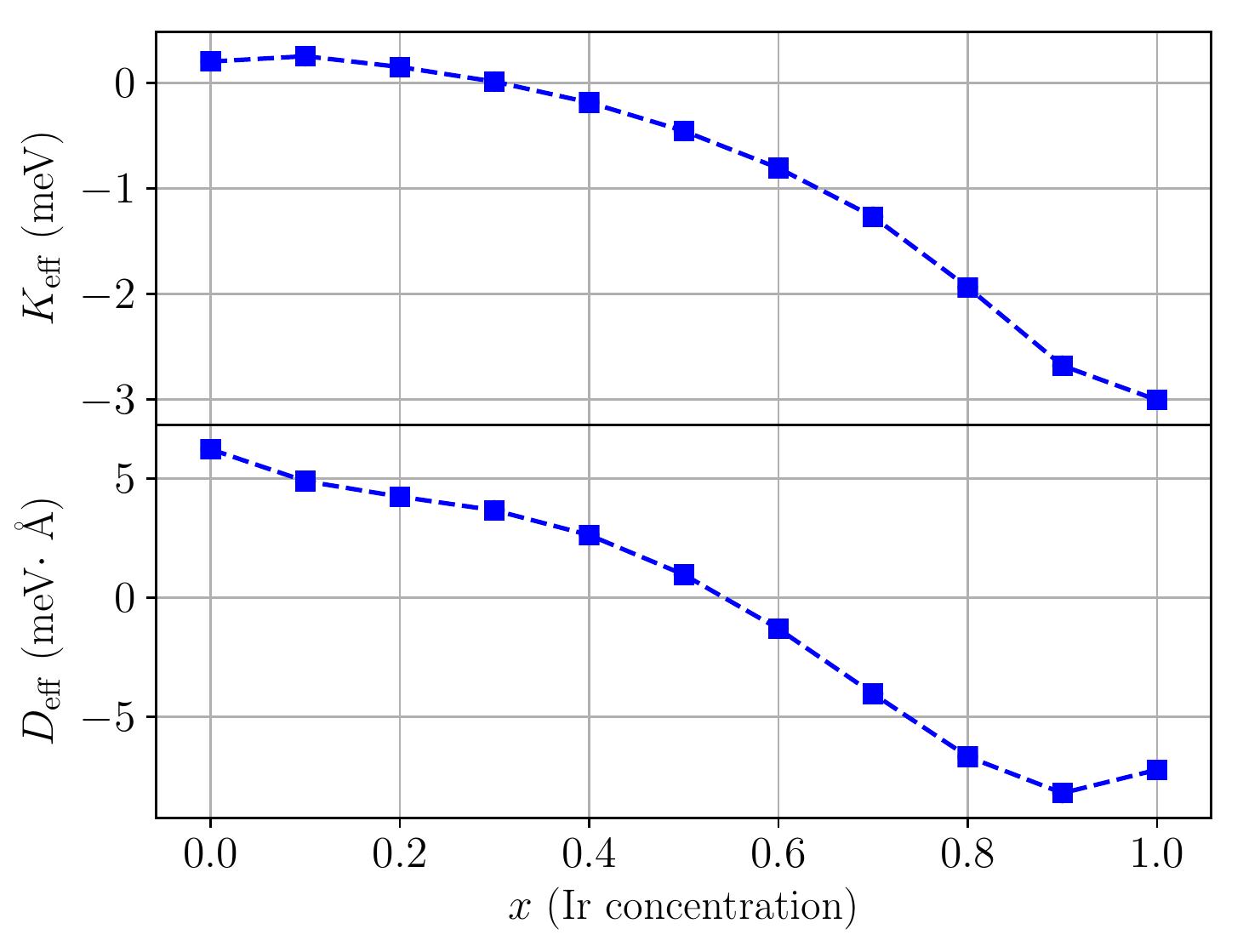}\\
\caption{(Color online) Calculated total MAE $K_{\rm eff}$ and effective DMI $D_{\rm eff}$ in the Co monolayer as a function of Ir concentration ($x$) in the Au$_{1-x}$Ir$_{x}$/Co/Pt(111) system.}
\label{fig:alloying}
\end{figure}

The changes of the effective DMI and MAE are shown in Fig.~\ref{fig:alloying} as a function of the Ir concentration. Unlike the case where the SOC was scaled (Fig.~\ref{eff-scaling}), the variation of $K_{\rm eff}$ and $D_{\rm eff}$ with $x$ is nonmonotonic, with a maximum of $K_{\textrm{eff}}$ at around $10$\% and a minimum of $D_{\textrm{eff}}$ at around $90$\% Ir concentration.

\subsection{Magnetic ground states\label{sec4}}

The ground states of the systems were determined by combining harmonic spin spiral calculations with spin dynamics simulations as described in Sec.~\ref{sec2c}. After scaling out the energy and length scales, the micromagnetic energy density in Eq.~(\ref{free-energy-mmm}) can be described by a single dimensionless parameter $\mathcal{K}\mathcal{J}/\mathcal{D}^{2}$, which governs the formation of the magnetic ground state. As already discussed in earlier publications \cite{PhysRevB.78.140403,PhysRevB.93.020404,PhysRevB.96.014423}, noncollinear ground states are expected to be formed for $-1<\mathcal{K}\mathcal{J}/\mathcal{D}^{2}<\frac{\pi^{2}}{16}\approx 0.62$ in this model; the upper limit denotes where magnetic domain walls become energetically favorable in out-of-plane oriented ferromagnets, while the lower limit indicates the instability of the in-plane oriented ferromagnetic state towards the formation of an elliptic conical state.

The calculated values are summarized in Table~\ref{table-gs} for these systems. For most considered capping layers the parameter $\mathcal{K}\mathcal{J}/\mathcal{D}^{2}$ is outside the range where the formation of noncollinear states is expected, and in the simulations we indeed observed ferromagnetic (FM) ground states. This can be explained either by the strong easy-plane (Os) or easy-axis (Ir) anisotropies, the weakness of the DMI for the Pt/Co/Pt(111) system, or the combination of the above for the Au capping layer. We note that since the micromagnetic model is isotropic in the plane, no long-range order is expected in in-plane ferromagnetic systems at finite temperature; however, this degeneracy may be lifted by weaker interactions not included in Eq.~(\ref{free-energy-mmm}), such as sixth-order anisotropy.
\begin{table}[t!]
\centering
\begin{ruledtabular}
\begin{tabular}{l r r}
     & $\mathcal{K}\mathcal{J}/\mathcal{D}^{2}$ & ground state    \\
\hline
Re   &  -0.20  &     tilted SS\\
Os &  -6.77   &      in-plane FM                  \\
Ir &  2.27   &       out-of-plane FM                    \\
Pt &  436.13   &      out-of-plane FM                                \\
Au &  -1.46   &     in-plane FM                                     
%overlayer     & $D/J$ & $K/J$    \\
%\hline
%Re   &  2.22  &     2.89\\
%Os &  0.11   &      0.26                  \\
%Ir&  -0.25   &       -0.43                    \\
%Pt &  0.02   &      -0.03                                \\
%Au &  0.13   &      0.004                                      
\end{tabular}
\end{ruledtabular}
\caption{Obtained magnetic ground states for X/Co/Pt(111) thin films (X=Re, Os, Ir, Pt, Au).}
\label{table-gs}
\end{table}
For the Re/Co/Pt(111) system, the micromagnetic model predicts \cite{PhysRevB.93.020404,PhysRevB.96.014423} a cycloidal spin spiral ground state with the normal vector in the plane, just as it was assumed in Eq.~(\ref{eqn4}). However, by minimizing Eq.~(\ref{eqn1}) with respect to the wave vector $\vec{k}$ and the normal vector $\vec{n}$, we obtained a tilted spin spiral state of the form
\begin{align}
\vec{s}_{i}=\vec{e}_{x}\cos k R_{i}^{y}\sin\Phi_{0}-\vec{e}_{y}\sin k R_{i}^{y}+\vec{e}_{z}\cos k R_{i}^{y}\cos\Phi_{0}.\label{eqntiltedSS}
\end{align}

\begin{figure}[b!]
\centering
\includegraphics[width=\columnwidth]{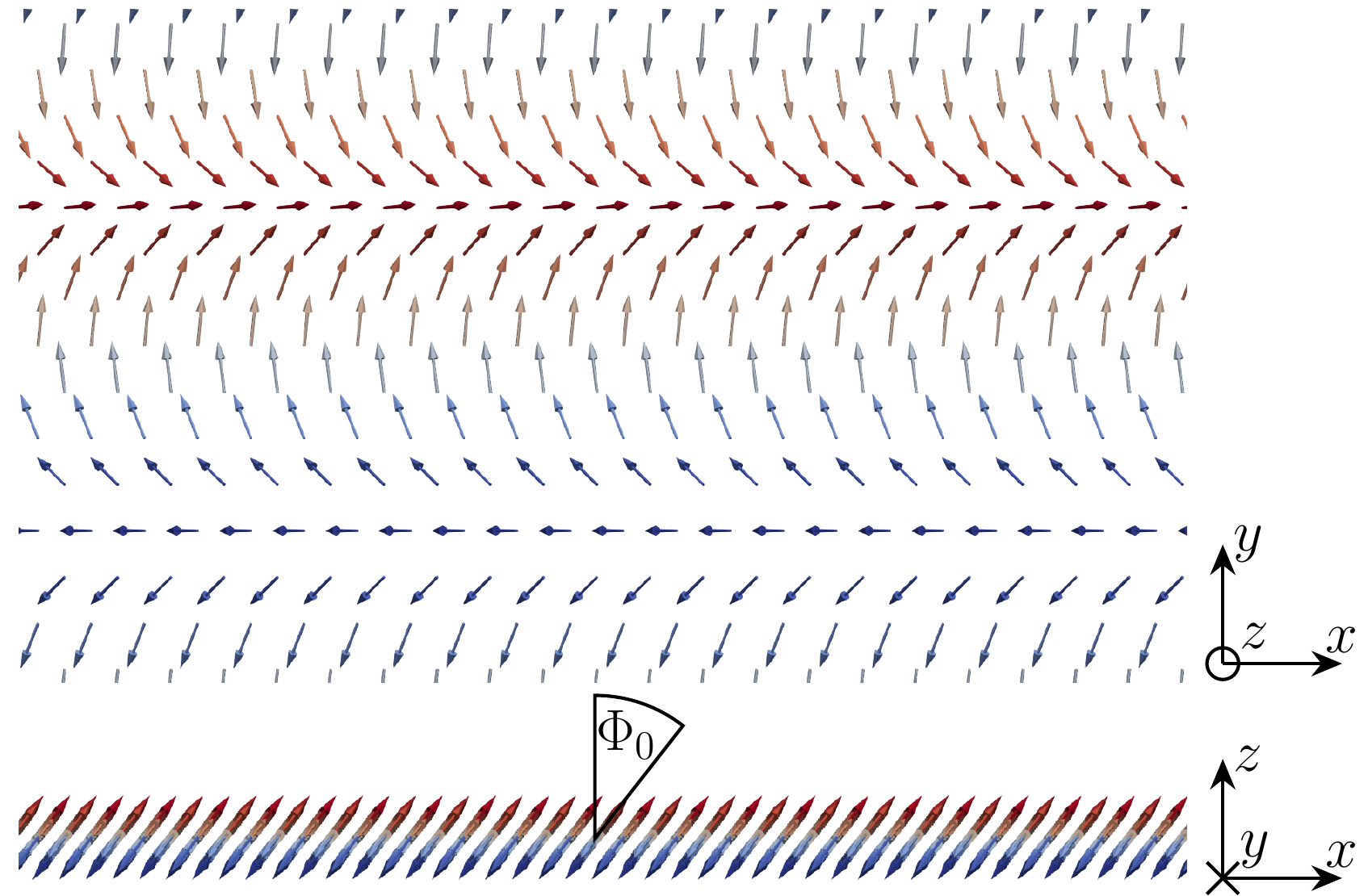}
\caption{The tilted spin spiral ground state found in the Re/Co/Pt(111) system in spin dynamics simulations. The tilting angle $\Phi_{0}$ is defined in Eq.~(\ref{eqntiltedSS}). Red and blue colors correspond to positive and negative out-of-plane spin components, respectively.}\label{fig-tiltedspiral}
\end{figure}

The ground state obtained from the spin dynamics simulations is displayed in Fig.~\ref{fig-tiltedspiral}. Although the spiral became slightly distorted due to the anisotropy, we found that it could still be relatively well described by Eq.~(\ref{eqntiltedSS}) using a wavelength of $\lambda=2\pi/k\approx3.5\,\textrm{nm}$ and a tilting angle of $\Phi_{0}\approx 38^{\circ}$. Note that the tilted spin spiral state is still a cycloidal spiral in the sense that the wave vector is located in the rotational plane of the spirals, but the normal vector is no longer confined to the surface plane. This is different from the case of weak DMI in out-of-plane magnetized films, where the normal vector of domain walls gradually rotates in the surface plane from N\'{e}el-type to Bloch-type rotation due to the presence of the magnetostatic dipolar interaction -- see e.g. Ref.~\cite{Thiaville-DW}. It also differs from the elliptic conical spin spirals discussed in Refs.~\cite{PhysRevB.93.020404,PhysRevB.96.014423} because the tilted spin spiral state has no net magnetization. The formation of such a ground state can be explained by the easy-plane anisotropy preferring an in-plane orientation of the spiral, the DMI preferring a spiral plane perpendicular to the surface, and the simultaneous presence of competing ferromagnetic and antiferromagnetic isotropic exchange interactions in the system, the latter also leading to the reduced value of the effective $J_{\textrm{eff}}$ parameter for the Re capping layer in Table~\ref{table-iso}.

\section{Summary and Conclusions\label{sec5}}
In conclusion, we examined the X/Co/Pt(111) (X = Re, Os, Ir, Pt, Au) ultrathin films using first-principles and spin-model calculations. We determined the Co-Co magnetic exchange interaction tensors between different pairs of neighbors and the magnetic anisotropies. From the results of the \textit{ab initio} calculations we also determined effective and micromagnetic spin-model parameters for the Co layers. For the isotropic exchange couplings we found dominant ferromagnetic nearest-neighbor interactions for all systems, which decrease with the d-band filling of the capping layer. This effect due to the hybridization between the $3d$ states of the Co layer and the $5d$ states of the capping layer can be qualitatively explained within a Stoner picture, which also accounts for the similarly decreasing magnetic moment. Considering the effective isotropic couplings of Co,
we found significantly lower values for Re and Ir overlayers than what would be expected simply based on the decrease of the nearest-neighbor interactions; this we attributed to competing antiferromagnetic couplings with further neighbors.

We also investigated the in-plane Dzyaloshinskii-Moriya interactions of Co, and found it to be weaker for all capping layers compared to the uncapped Co/Pt(111) system. For the Ir capping layer we found a switching from counterclockwise to clockwise rotation, which is unexpected since the same switching can also be observed if the Ir is inserted between the magnetic layer and the substrate \cite{PhysRevB.94.214422}. We attributed this effect to the reduced coordination number of Ir atoms and the electrostatic potential barrier at the surface. We also found a correlation between the effective Dzyaloshinskii-Moriya interactions $D_{\rm eff}$, the effective magnetic anisotropies $K_{\rm eff}$, and the anisotropy of the orbital moment $\Delta m_{\textrm{orb}}$.
We further investigated the sign change of the effective Dzyaloshinskii-Moriya interaction of Co for the Ir capping layer by scaling the strength of the spin--orbit coupling at the Ir sites and by tuning the filling of the $5d$-band in a Au$_{1-x}$Ir$_{x}$/Co/Pt(111) system. We found a linear dependence of the effective Dzyaloshinskii-Moriya interaction on the spin--orbit coupling strength in agreement with the perturbative description, and a nonmonotonic dependence on the band filling.

Using the spin-model parameters we determined the magnetic ground state for all considered systems. For Os, Ir, Pt and Au capping layers we found a ferromagnetic ground state, in agreement with the analytical prediction based on the calculated micromagnetic parameters. For the Re/Co/Pt(111) system we found a tilted spin spiral ground state, the appearance of which can only be explained if competing ferromagnetic and antiferromagnetic isotropic exchange interactions are taken into account alongside with the Dzyaloshinskii--Moriya interaction and the easy-plane anisotropy.

Our results highlight the importance of \textit{ab initio} calculations and atomic spin-model simulations in cases where simpler model descriptions might lead to incomplete conclusions. The present study may motivate further experimental investigations in this direction, exploring the sign of the Dzyaloshinskii--Moriya interaction and the role of competing isotropic exchange interactions in ultrathin film systems.

\begin{acknowledgments}

The authors would like to thank F. Kloodt-Twesten and M. Mruczkiewicz for insightful discussions. Financial support of the National Research, Development, and Innovation Office of Hungary under Projects No.~K115575, PD120917 and FK124100, of the Alexander von Humboldt Foundation, of the Deutsche Forschungsgemeinschaft via SFB668, of the SASPRO Fellowship of the Slovak Academy of Sciences (project No.~1239/02/01), and of the Hungarian State E\"otv\"os Fellowship is gratefully acknowledged.

\end{acknowledgments}

\bibliographystyle{apsrev4-1}
\bibliography{5dCoPt111}
\end{document}